# ADVANCED SURFACE TREATMENTS FOR MEDIUM-VELOCITY SUPERCONDUCTING CAVITIES FOR HIGH-ACCELERATING GRADIENT CONTINUOUS-WAVE OPERATION

K. McGee[1,2], S. Kim[1], K. Elliott[1], A. Ganshyn[1], W. Hartung[1], P. Ostroumov[1], A. Taylor[1], T. Xu[1]
M. Martinello[2], G. V. Eremeev[2], A. Netepenko[2], F. Furuta[2], O. Melnychuk[2]
M. P. Kelly[3], B. Guilfoyle[3], T. Reid[3]

[1]Facility for Rare Isotope Beams, Michigan State University, East Lansing, MI, 48824, USA.
[2]Fermilab National Laboratory, Batavia, IL, 60510, USA.
[3]Argonne National Laboratory, Argonne, IL, 60439, USA.

Nitrogen-doping and furnace-baking are advanced high-$Q_0$ recipes developed for 1.3 GHz TESLA-type cavities. These treatments will significantly benefit the high-$Q_0$ linear accelerator community if they can be successfully adapted to different cavity styles and frequencies. Strong frequency- and geometry- dependence of these recipes makes the technology transfer amongst different cavity styles and frequencies far from straightforward, and requires rigorous study. Upcoming high-$Q_0$ continuous-wave linear accelerator projects, such as the proposed Michigan State University Facility for Rare Isotope Beam Energy Upgrade, and the underway Fermilab's Proton Improvement Plan-II, could benefit enormously from adapting these techniques to their $\beta_{opt} = 0.6$ ~650 MHz 5-cell elliptical superconducting rf cavities, operating at an accelerating gradient of around ~17 MV/m. This is the first investigation of the adaptation of nitrogen doping and medium temperature furnace baking to prototype 644 MHz $\beta_{opt} = 0.65$ cavities, with the aim of demonstrating the high-$Q_0$ potential of these recipes in these novel cavities for future optimization as part of the FRIB400 project R&D. We find that nitrogen-doping delivers superior $Q_0$, despite the sub-GHz operating frequency of these cavities, but is sensitive to the post-doping electropolishing removal step and experiences elevated residual resistance. Medium temperature furnace baking delivers reasonable performance with decreased residual resistance compared to the nitrogen doped cavity, but may require further recipe refinement. The gradient requirement for the FRIB400 upgrade project is comfortably achieved by both recipes.

## I. INTRODUCTION

High-$Q_0$ superconducting rf cavity development plays a critical role in evolving the state of the art in nuclear and high-energy physics experimentation. Michigan State University's Facility for Rare Isotope Beams (MSU/FRIB) is a superconducting (SC) heavy-ion linac supporting world-leading nuclear physics research. The FRIB baseline design achieved fully commissioned status in January 2022, and is now providing beam to user experiments [1]. The FRIB400 energy upgrade [2] proposes to double the beam energy from 200 to 400 MeV/u for the heaviest Uranium ions, or about 1 GeV for protons. This upgrade would significantly evolve the field of rare isotope science, broadening the scope of FRIB's role in the nuclear physics research community to address current, critical questions in the fields of nuclear structure, nuclear astrophysics, tests of fundamental symmetries, and industrial applications of isotopes [2].

A design study [3] for the FRIB400 upgrade identified $\beta_{opt} = 0.65$ 644 MHz 5-cell elliptical SC niobium rf cavities as the only design capable of providing the needed beam energy gain within the 80 m of space available in the FRIB tunnel (Table 1, Figure 1). This proposed 5-

cell cavity operating at medium gradient bridges a hitherto unfilled gap between low-β TEM-type cavities and β = 1 elliptical-type cavities for continuous wave (CW) operation.

Table 1: FRIB400 Cavity Parameters

| | |
|---|---|
| Frequency | 644 MHz |
| Geometric β | 0.61 |
| Optimal β | 0.65 |
| Aperture diameter | 83 mm |
| Effective length $L_{eff}$ | 71.0 cm |
| Number of cells | 5 |
| Geometric shunt impedance $R/Q$ | 368 Ω |
| Geometry factor $G$ | 188 Ω |
| $E_{peak}/E_{acc}$ | 2.28 |
| $B_{peak}/E_{acc}$ | 4.42 mT/(MV/m) |
| **2 K operating goals** | |
| Accelerating gradient $E_{acc}$ | 17.5 MV/m |
| Peak surface electric field $E_{peak}$ | 40 MV/m |
| Peak surface magnetic field $B_{peak}$ | 77.5 mT |
| Intrinsic quality factor $Q_0$ | $2 \times 10^{10}$ |

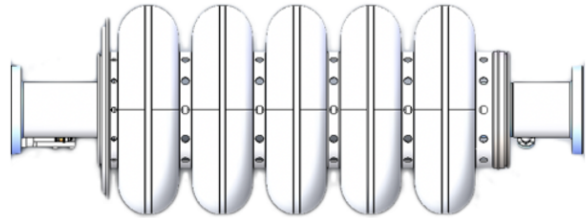

Figure 1: Drawing of the FRIB400 5-cell cavity.

As a continuous-wave (CW) linac, FRIB needs field-leading high-$Q_0$ superconducting resonators to minimize losses and fit the upgraded segment, 55 cavities in 11 cryomodules, within a small upgrade to existing cryomodule facilities. In parallel, the baseline design for the Proton Improvement Plan II (PIP-II) includes a segment of 36 $β_{opt}$ = 0.65 650 MHz 5-cell elliptical SC niobium rf cavities compatible with CW operation [4]. The broad similarities between these two designs make the investigation herein relevant to both, despite being conducted entirely on FRIB400 644 MHz cavity prototypes.

The Spallation Neutron Source (SNS) and European Spallation Source (ESS) both employ buffered chemical polished sub-GHz "medium-β" cavity designs similar to FRIB400, however as pulsed machines, their $Q_0$ requirements are set around an order of magnitude lower than that of the FRIB400 cavities, and, in the production runs for each machine, buffered chemical polishing did not consistently reach FRIB400's minimum required $Q_0$ of $2 \times 10^{10}$ at an accelerating gradient of 17.5 MV/m [5,6].

Previous work [7] validated the FRIB400's novel $\beta_{opt}$ = 0.65 5-cell elliptical 644 MHz geometry. Using standard RF processing techniques, both prototype cavities achieved acceptable $Q_0$ performance near $2\times10^{10}$, meeting the FRIB400 minimum design goal, and showed that cavity multipacting was completely conditionable in about 30 min. The steep side-walls of this cavity design had lead to some concern over its susceptibility to the "accordion" mechanical mode around 94 Hz, however measurement of the mode in the prototype cavity indicated the quality factor of this mechanical mode is no higher than that in currently operating 1.3 GHz TESLA cavities or FRIB 322 MHz half-wave resonators [7]. The EP-only treatment was found to produce the highest $Q_0$ of $2.3\times10^{10}$ at 17.5 MV/m, compared to BCP treatments, and applications of low-temperature (120°C) baking.

A vertical-test result of a $Q_0$ of $2.3 \times 10^{10}$ at 17.5 MV/m satisfies the FRIB400's minimum $Q_0$, however it is prudent to anticipate degradation in $Q_0$ from vertical cavity test to cryomodule installation, given the more complex operational environment of cavities experience once installed in cryomodules. The large size and steep sidewalls of these cavities make cleaning and handling these cavities particularly challenging, increasing the likelihood of $Q_0$ degradation from vertical-test to cryomodule installation. Thus, a comfortable margin over the $Q_0$ reported in [7] is highly desirable in view of the ultimate production of these cavities for projects like FRIB400 and PIP-II. Moreover, increasing the cavity $Q_0$ by a factor of two or more from the baseline requirements would lead to substantial operational and project cost savings for these machines.

The success of N-doping in 1.3 GHz TESLA cavities employed in LCLS-II highlights the potential for significant operational cost savings if N-doping can be successfully adapted to a different frequency and velocity regime. Likewise, current and ongoing efforts show promising results for furnace baking: for example, CEPC recently demonstrated impressive success with a 650 MHz single-cell $\beta$ = 1 SRF cavity which achieved a $Q_0$ of $6.4 \times 10^{10}$ at 30 MV/m [8] [9].

This adaptation however is far from straightforward, and in the context of the FRIB400 upgrade proposal, requires specific investigation and optimization before the technology can be proposed for industrialization for medium-velocity 644 MHz cavities. The N-doping effect has known frequency dependence, and the N-doping process is complex and sensitive enough to cavity geometry that modifications from the 1.3 GHz process need to be made to accommodate cavities like the FRIB400 prototype. These modifications, including significant alterations to the EP process, and the sheer challenge of treating and handling a large cavity in the context of volume- and flow- dependent processes ultimately mean that while 1.3 GHz cavities serve as a good proof of principle, focused study is required to transfer the technology to different frequency ranges. Cryogenic cost savings on this scale, in addition to improved $Q_0$ design goal margins, highly motivate departure from conventional EP and thus this investigation into methods of effectively adapting nitrogen-doping and furnace-baking to the novel $\beta_{opt}$ = 0.65 5-cell elliptical 644 MHz geometry.

## II. ADVANCED TREATMENTS

### II.1 Nitrogen Doping

Nitrogen doping (N-doping) is an advanced, record $Q_0$-achieving treatment developed at Fermi National Accelerator Laboratory (FNAL) in 2013 [10,11]. In typical N-doping processes, a bulk-electropolished (EP) niobium (Nb) cavity is baked in a vacuum furnace at 800°C with nitrogen gas introduced at 25 mTorr for some minutes. The cavity is then annealed without nitrogen for some minutes, then allowed to cool before undergoing light EP removal of 5-7 μm. This has produced field-leading high-$Q_0$ and high-gradient results in 1.3 GHz TESLA cavities, and on the strength of these results, N-doping saw rapid industrialization for the delivery of the Linear Coherent Light Source II (LCLS-II) project [11-13].

However, key physical properties of the N-doping effect make its translation to a "low" (sub-GHz) frequency, medium-field application challenging. The cavity's quality factor is an inverse function of the superconducting surface resistance, $R_S$, which is generally broken into the temperature-dependent (commonly, $R_{BCS}$) and temperature independent ($R_0$) portions:

$$R_S = R_{BCS}(T) + R_0 \qquad (1)$$

where the temperature-dependent portion in the simple case of low fields generally follows from the Bardeen-Cooper-Schrieffer theory of superconductivity, :

$$R_{BCS} = \frac{A\omega^2}{T} e^{-\frac{\Delta}{kT}} \qquad (2)$$

A numerical fit of (2) in the lower field, clean limit (3) [14] estimates $R_{BCS}$ of niobium as a function of operating temperature (T) and operating frequency (f, GHz):

$$R_{BCS}(\Omega) = 2 \times 10^{-4} \frac{1}{T}\left(\frac{f}{1.5}\right)^2 e^{-\frac{17.67}{T}} \qquad (3)$$

This model holds well for the FRIB400 EP cavities: the calculated value at 2 K is approximately 2.68 nΩ, and in the EP FRIB400 cavities, the measured values are around 2.7 nΩ. The field-dependence of $R_{BCS}$ is less well-modeled [15], however, and is of particular interest.

In EP cavities, $R_{BCS}$ is the dominant contribution to $R_s$, and increases as a function of field: we found previously the ratio of $R_{BCS}$ to $R_0$ grows from roughly 2 at 2 MV/m to 2.4 at 17.5 MV/m in FRIB400 cavities ($B_{pk}$ 77 mT) [7]. Martinello et al [16] found that the ratio of $R_{BCS}$ to $R_0$ after EP was approximately 3 in high-β 650 MHz PIP-II cavities at similar peak fields. Moreover, previous work by Martinello et al. [17,18] showed that the cavity operating frequency strongly affects the strength of this field-dependence of $R_{BCS}$: in the medium-field range, $R_{BCS}$ tends to decrease as a function of field at frequencies higher than ~1 GHz in n-doped cavities. However, as the operating frequency lowers, this behavior gradually reverts back to the more typical increase in $R_{BCS}$ as a function of field.

Low-field $R_{BCS}$ is decreased by N-doping as compared to EP for all frequencies, and the magnitude of this decrease grows with operating frequency [18]. The authors note it is plausible that the frequency dependence of $R_{BCS}$ results from the applied rf field accessing frequency-dependent nonlinear electrodynamic effects within the superconducting niobium [18]. As demonstrated in aluminum resonators in [19], applied microwave fields can support a frequency-dependent non-equilibrium distribution of quasiparticle states in superconductors. The resultant broadening of the density of quasiparticle states decreases the population of quasiparticles near the superconducting gap, making them less susceptible to losses due to photon absorption. Above minimum temperatures, [19] finds that this effect can lead to increasing $Q_0$ as a function of the applied field. Recent theoretical work in the dirty limit of niobium has confirmed one can expect the amplitude of these quasiparticle states, and thus the nonlinear effects, to be significantly attenuated below operating frequencies of about 1 GHz [20].

While this suggests we cannot expect to reap the benefit of the anti-Q slope in 650 MHz cavities with current N-doping recipes, Martinello et al. [17,18] showed that N-doping still lowers the rate of $R_{BCS}$ increase at medium fields, such that one can estimate that at the FRIB400 peak magnetic field (Bpk) of 77 mT, $R_{BCS}$ is still likely to be lower overall after N-doping. The authors find that in both single and multi-cell 650 MHz cavities that N-doping delivered the best quality factors of the tested treatments, despite this detrimental frequency dependence [18]. These N-doped cavities experienced quench fields on average around 24 MV/m [17], which is comfortably above FRIB400 operating gradient, 17.5 MV/m.

An expected challenge in applying N-doping to the FRIB400 prototypes lies in the eccentric geometry of the medium-β cavities. Martinello et al. [16] found N-doped HB650 cavity performance was highly sensitive to post-doping EP removal to within a few μm. The large cavity aspect ratio, and thus large standoff distance of parts of the cavity surface from the EP electrode, raises concerns over our ability to achieve uniform surface removal on a small scale. Previous work [7] presented a modified EP cathode for the FRIB400 medium-β cavities, and validated bulk EP removal uniformity over ~150 μm.

## II.2 Furnace Baking

A second rf surface treatment of interest, furnace baking, in which Nb cavities are vacuum-furnace baked at moderate temperatures between 300-500 °C for a few hours, has recently achieved impressive quality factors approaching $5 \times 10^{10}$ in 1.3 GHz cavities. In some cases, furnace baking has even elicited the anti-Q slope seen in N-doped cavities [21]. The origin of the furnace baking effect has been recently investigated from the standpoint of the dissolution and diffusion of oxygen from the native oxide layer on the niobium surface into the bulk Nb [22-24]. Interstitial oxygen may play a similar role to nitrogen in moving the superconductor towards the dirty limit in which the nonequilibrium superconducting effects discussed previously may be accessed. Furthermore, analogous to what has been suggested in the case of N-doping, in which the interstitial nitrogen appears to interrupt the formation of lossy niobium-hydrides [10], Wenskat et al. [24] find that with the diffusion of oxygen into the bulk, hydrogen-vacancy complexes, which function as nucleation sites for lossy niobium hydrides, are preferentially replaced by oxygen-vacancy complexes. These in turn prevent the formation of niobium hydrides, and reduce the inherent residual resistance of the cavities. Further evidence suggests that the surface layers of niobium oxides are structurally reordered during the furnace baking, which also may contribute to reduced temperature-independent rf losses [24].

Furnace baking is distinct from another recently developed treatment in this temperature range called "Mid-T" baking [25], in which the cavity is baked *in situ*. Unlike furnace baking, this *in-situ* method avoids cavity air exposure post-bake, preventing the native oxide layer from re-forming before RF testing. While very good results have been achieved [25], it is currently unclear exactly how detrimental this reformed oxide layer is to the ultimate rf performance, which must be weighed against the added complexity of performing *in-situ* baking. Due to our own interest in immediate accelerator applications of this technique, we chose to investigate furnace baking as the more practically achievable process in the FRIB400 prototype cavities.

In summary, the goal of this work is to analyze the potential of N-doping and furnace baking to maximize $Q_0$ in FRIB400's novel 644 MHz $\beta_{opt}$ = 0.65 cavities. The most promising of these treatments will emerge from this study as a strong candidate for future systematic development as the proposed RF treatment process to be applied to the 55 FRIB400 cavities. Due to close similarities in the two designs, this work will also usefully inform PIP-II 5-cell elliptical "low-beta" ($\beta_{opt}$ = 0.61) 650 MHz CW cavity research development (LB650).

## III. RF SURFACE PROCESSING AND CAVITY PERFORMANCE

Two FRIB400 prototype cavities, S65-001 and S65-002, were fabricated by Research Instruments GmbH and used previously [7] to study the "conventional" RF surface treatments at Michigan State University/FRIB, Electropolishing, BCP, and low-temperature baking. Afterwards, both cavities were reset with at least 50 μm EP, the last 10 μm of which being "cold EP," and baseline $Q_0$ vs $E_{acc}$ measurements were retaken. The EP results were consistent with previous findings after this "reset," demonstrating that the effect of the previous series of cavity treatments had been successfully removed. Both cavities underwent hydrogen degassing [7] and no other high-temperature treatments were applied to these cavities prior to this study. All EP processing steps were performed at the EP facility at Argonne National Laboratory, using the modified EP cathode described previously [7].

N-doping of rf cavities canonically consists of heating the cavity to 800°C in a vacuum-furnace, then introducing 25 mTorr of nitrogen gas for some minutes, followed by some minutes of annealing at 800°C, before cooling the cavity to room temperature. Martinello et al. [16] studied 2 min doping / 6 min annealing (2/6) and 3 min doping / 60 min annealing (3/60) recipes and found 2/6 delivered superior cavity performance. Subsequent N-doping study in 1.3 GHz cavities suggested that 2/0 doping had lower sensitivity to trapped magnetic flux [26]. This recipe with a post-doping "cold" EP [27] was thus chosen and refined in 1.3 GHz cavities for the LCLS-II HE verification cryomodule with good results [26]. In this study, we elected to investigate both 2/6 and 2/0 N-doping treatments with "cold" post-doping EP for the FRIB400 cavities. Martinello et al. found that 7 μm was optimal in 2/6 N-doped high-β 650 MHz cavities [16], which we adopted for the $\beta_{opt}$ = 0.65 644 MHz cavities.

The modified EP cathode [7] consists of aluminum "doughnuts" alternating with Teflon masking, to increase the radius of the cathode in the equator region and attenuate the field in the iris region. A recently-fabricated single-cell version of the FRIB400 cavity allowed the opportunity to take ultrasonic thickness measurements at many intervals across the cavity surface to check the uniformity of bulk EP removal with the new cathode design (Figure 2). We verified that the bulk EP of 150 μm EP + 10 μm "cold" EP achieved 160 μm removal with a standard deviation across the entire cavity cell surface of approximately 6 μm. This result affirmed that the eccentric aspect ratio of the $\beta_{opt}$ = 0.65 cavities does not appear to pose insurmountable challenges to EP processes.

At the Argonne EP facility, the cavity surface temperature during EP is monitored with 8 thermocouples placed as follows: one on each beam tube, at each equator, and one additional sensor on a cavity iris. In the 1-cell case, the 8 thermocouples are placed on each beam tube, on each iris, halfway up each sidewall, and on either side of the equator weld. These sensors monitor the temperature throughout the EP process, and show the average temperature of the EP process is stably controlled within its natural cyclical variation of approximately +/- 1.5 °C of the target temperature, as the monitors rotate into and out of the acid bath (e.g., between 15-17 °C for "cold EP"). The maximum temperature of this oscillation varies by under half a degree per cycle, demonstrating stable temperature control throughout the process.

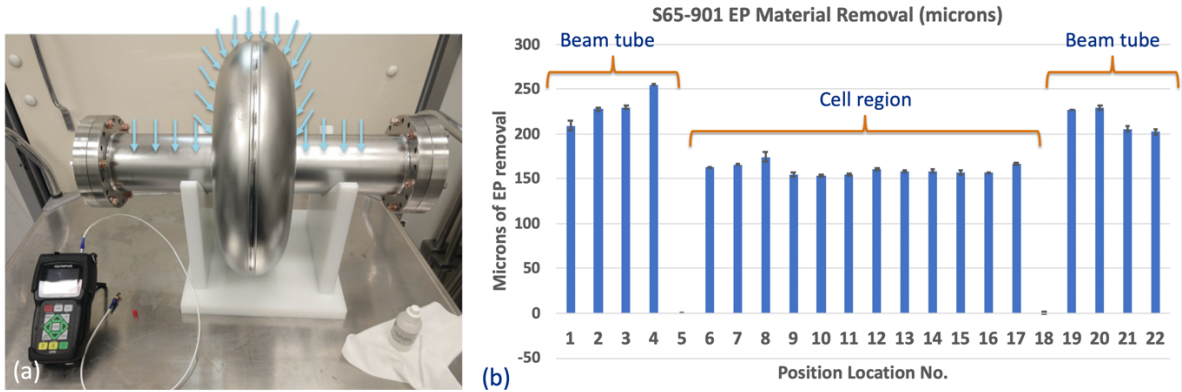

Figure 2. a) Image of the prototype FRIB 644 MHz $\beta_{opt}$ = 0.65 single cell cavity, with blue arrows indicating the 20 ultrasonic thickness measurement locations across the cavity surface. b) Microns (μm) of removal, with measurement error, showing an average removal in the cell region of 160 μm with a standard deviation of approximately 6 μm. Note no systematic bias exists between the material removal from the near-vertical sidewalls (positions 6-10, 13-17) and the cavity equator (position 11-12). No measurements could be taken at positions 5 and 18 due to the high curvature of the iris in that region, and thus were left blank in the image.

RF testing was conducted at both MSU and FNAL vertical test facilities using the measurement and error calculation methods that have been previously described in [28]. Both have implemented background magnetic field cancellation in addition to standard shielding: MSU cavity test stand fits several turns of wire connected to a DC power supply around the opening of the magnetic shielding vessel in the cavity test Dewar. The current in the wire is adjusted until a fluxgate magnetic probe mounted in the Dewar midway down on the cavity surface reads ~1mG or less of magnetic field. A magnetic survey of the Dewar interior was performed, in which magnetic fluxgate probes were lowered in intervals, and ambient field measurements were taken at the cavity equator radius with the cavity installed in the Dewar, and the magnetic shielding in place. These measurements demonstrate that the combination of magnetic shielding and field cancellation coils installed is effective at minimizing background magnetic field for the length of the cavity (Figure 3a). The FNAL vertical test stand uses active magnetic field cancellation in the form of Helmholtz coils mounted within the Dewar on the cavity fixture, maintained in a control loop (Figure 3b). This also generally achieves 1mG or less of background magnetic field during cooldown.

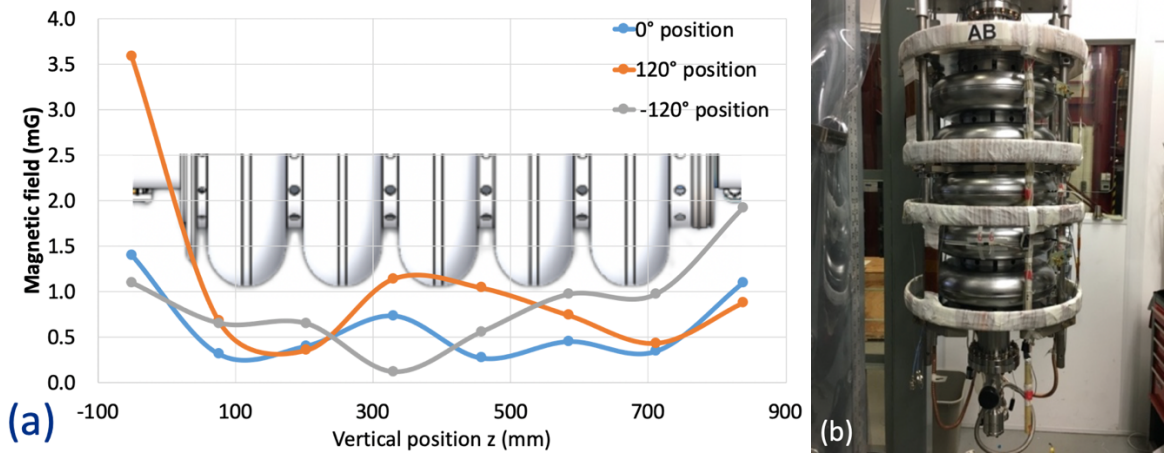

Figure 3. 3a) Magnetic survey of the MSU/FRIB vertical test stand (VTS) Dewar demonstrating efficacy of the MSU/FRIB field cancellation technique. A cartoon-cavity is overlaid to show approximately where the cavity equators lay in relation to the vertically scanned magnetic field in the Dewar at the three 120-degree offset positions. 3b) S65-001 644 MHz FRIB400 cavity after clean assembly to the vertical test insert at FNAL. The installed Helmholtz coils dynamically compensate the measured background magnetic field during cooldown, so field applied to the cavity is constrained to 1 mG or less for most tests. In both cases, the niobium cavity was electrically isolated from the titanium cavity fixture (support frame).

After fast cooldowns in the compensated magnetic field, the cavities were maintained at 2K and the initial $Q_0$ vs $E_{acc}$ curves were taken (Figure 4). At the FNAL facility, if the cavities experienced quench, as happened in the N-doped cases, the 2 K Q-curve was retaken to benchmark the effect of the trapped flux that may have entered the cavity during the quench.

The FNAL vertical test stand (VTS) facility has the ability to achieve temperatures of 1.5 K or lower in the test dewar. The rf surface resistance ($R_s$) consists of temperature-dependent and temperature-independent portions, and the ability of the FNAL VTS to achieve 1.5 K or less in the dewar allows us to nearly completely eliminate the temperature-dependent portion of the RF surface resistance. The details of this method can be found in [29]. Throughout this analysis, the cavity surface resistance is measured as the geometry factor divided by the quality factor.

At 1.5 K, the measured surface resistance is essentially equal to the temperature-independent portion of the RF surface resistance, $R_0$. This temperature-independent portion includes resistance arising from the cavity's sensitivity to trapped magnetic flux, as well as that arising from intrinsic material imperfections in the niobium material. The portion of $R_0$ arising from the trapped magnetic flux changes as a function of the quantity of trapped magnetic flux, however, for all tests in this paper, field cancellation methods were employed to minimize the background field to ~1mG or less. The $R_0$ reported thus represents both trapped flux resistance and intrinsic resistance from niobium material imperfections. We regard it as a useful measure of the $R_0$ that may, best-case, be achieved in either VTS or cryomodule contexts, where both magnetic and intrinsic sources of $R_0$ are at play. The temperature-dependent $R_{BCS}$ is then calculated by subtracting the $R_0$ measured at low-temperature from the overall cavity surface resistance measured at 2 K.

This decomposition provides insight to the functionality of the rf treatments under investigation. As shown in Figure 5, nitrogen-doping lowers $R_{BCS}$, but does not produce the anti-Q slope characteristic of 1.3 GHz nitrogen-doped cavities. While N-doping often reduces $R_0$ in 1.3 GHz cavities, we find that $R_0$ at this frequency is increased. In medium-temperature baking, a more modest decrease in $R_{BCS}$ is observed, but $R_0$ is improved compared to the nitrogen-doped cavity. Prior to this study, the magnitude of these changes were not known, and could not be inferred from 1.3 GHz cavity data, given the known frequency-dependence of these effects.

Prior to the treatments studied in this paper, the 5-cell FRIB400 644 MHz prototype cavities S65-001 and S65-002 underwent 50 μm EP resets, with the last 10 μm cold EP. All EP treatments were conducted at 18 V. S65-001 was treated with an 800°C 2/6 N-doping followed by a 7 μm post-doping cold-EP. Subsequent rf tests at both MSU/FRIB facilities and at the Fermilab facilities showed increased residual resistance in this cavity, leading to a $Q_0$ lower than the EP baseline. This was despite similar background magnetic field levels, the implementation of fast-cooldowns, and the electrical isolation of the cavity from its fixture. We therefore suspected the post-doping EP had not been complete. The cavity then received an additional

5 μm EP, which decreased $R_0$ by nearly 1 nΩ at 17.5 MV/m, suggesting a more complete removal of the niobium nitride layer, particularly since the background magnetic field was slightly enhanced in this trial.

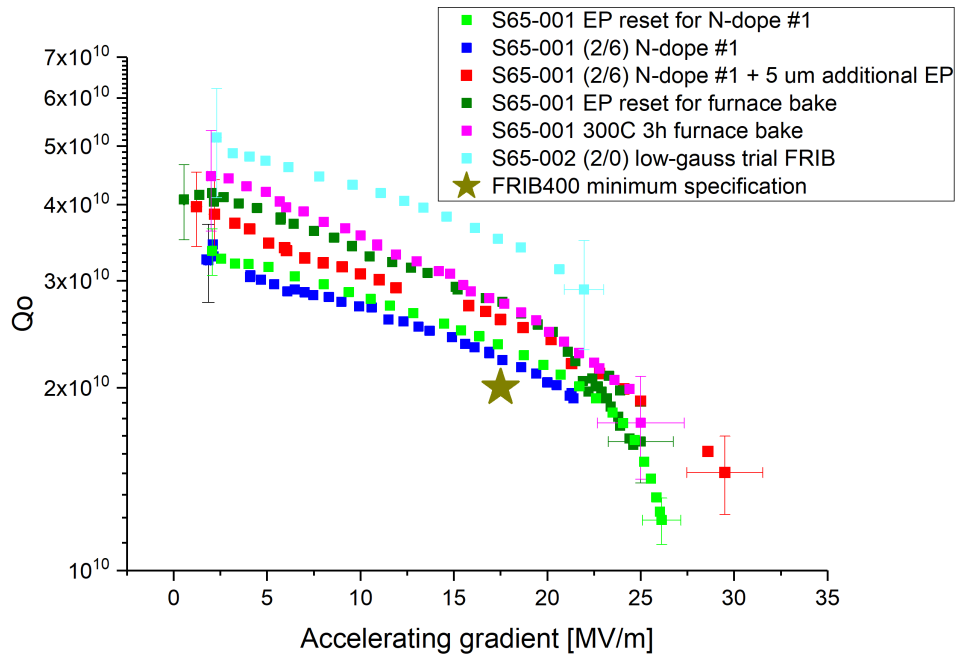

Figure 4. 2 K $Q_0$ vs. $E_{acc}$ curves for 2/6 doped, 2/0 doped, and 300°C 3h furnace baked FRIB 644 MHz $\beta_{opt}$ = 0.65 5-cell elliptical cavities. The first 2/6 doping of S65-001 slightly reduced performance from its EP baseline (green) by about 5% at the cavity's operating gradient of 17.5 MV/m. An additional 5 μm EP improved the N-doped $Q_0$ performance by 11% from the EP baseline (red). 2/0 doping of S65-002 delivered the best performance of all trials (cyan). The second EP reset of S65-001 saw an improvement over the previous EP baseline of over 20% (dark green). The 300°C 3-hour furnace bake that immediately followed this EP reset (magenta) produced only slight improvement over the 2$^{nd}$ EP reset. Post-EP tests were limited by available amplifier power; the N-doped tests were limited by quench; and the furnace baking test reached the administrative limit of 26 MV/m in the presence of mild field emission.

S65-001 then underwent a 60 μm EP reset, which resulted in a higher baseline EP performance than the first EP reset. To illustrate the comparison, 1$^{st}$ reset had 2.2 nΩ of $R_0$ at $E_{acc}$ = 2 MV/m (fitted from an $R_S$ vs. T measurement) [7], and the 2$^{nd}$ reset had approximately 1.7 nΩ of $R_0$ at $E_{acc}$ = 2 MV/m, leading to a difference in EP baseline $Q_0$ of nearly 27% at 2 MV/m, which became closer to 20% at higher fields. With a single-cell measured sensitivity of 0.45 nΩ/mG [30], this suggests there was about 1mG less trapped flux in the second trial, conducted at FNAL, which is possible given differing cooldown dynamics and not-yet-optimized field cancellation techniques employed at FRIB/MSU at the time.

The subsequent furnace baking treatment was straightforward: 300°C 3-hour bake in a high-vacuum furnace, followed by high-pressure water rinsing (HPR) and clean assembly to the test insert. While the furnace baking improved upon the 2/6 N-doped results, decreasing $R_0$ by 19% and only increasing $R_{BCS}$ by 7.5%, it did not significantly elevate the $Q_0$ performance above the immediately prior EP baseline performance. However in the breakdown of $R_{BCS}$ and $R_0$

(Figure 5), it is apparent that furnace baking decreased $R_{BCS}$ and increased $R_0$ in equal, opposite proportions compared to the EP baseline. This overall suggests that preference for N-doping or furnace baking would be dependent on the target operating temperature of the prospective system, with higher-temperature systems preferring N-doping, and lower-temperature systems preferring furnace baking.

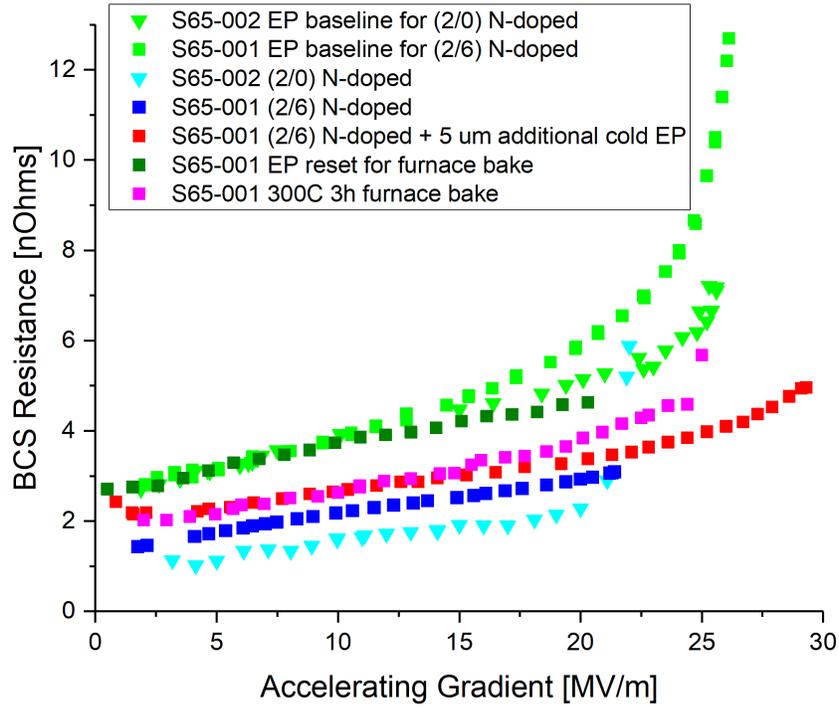

A)

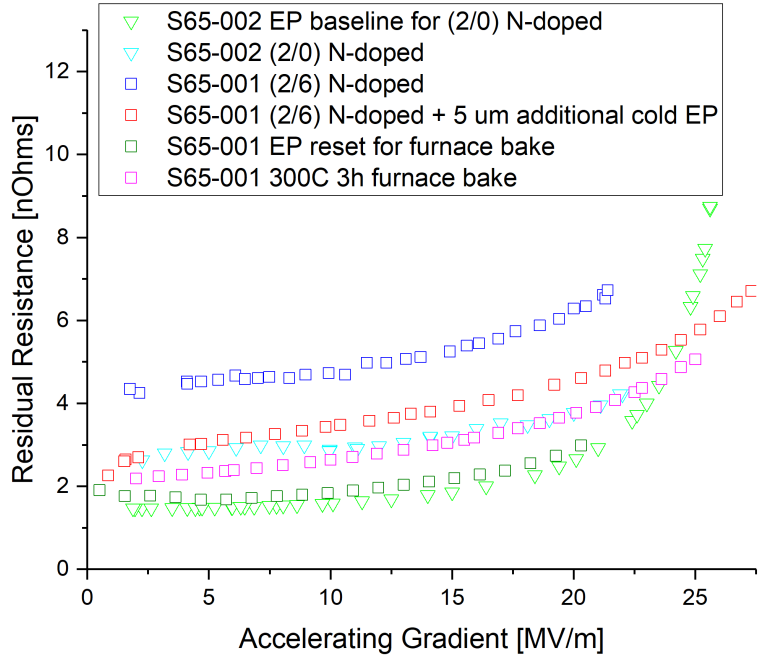

B)

Figure 5. A) Decomposed temperature-dependent surface resistance ($R_{BCS}$) and B) temperature-independent surface resistance ($R_0$, residual resistance) for both FRIB400 prototype cavities, S65-001 and S65-002. A). As expected, EP-treated cavities have the highest $R_{BCS}$ at medium field, and N-doped cavities have the lowest. Additional 5 μm EP in the 2/6 doped cavity (12 μm total post-dope) slightly increased the $R_{BCS}$ which is not consistent with [18] in which additional 10 μm EP lowered $R_{BCS}$ in 2/6-doped cavities. The medium-temperature baked cavity (teal) had moderate $R_{BCS}$, between that of the EP cavity and the N-doped cavity. B) Residual resistance was highest in N-doped cavities and lowest in EP cavities. The furnace baked treated cavity has higher $R_T$ than the N-doped cavity and lower R0 than the N-doped cavity.

The S65-002 EP baseline performance was essentially identical to the 2nd EP of S65-001, so is omitted from Figure 4 for clarity. The cavity then underwent 2/0 N-doping, and was tested at MSU/FRIB facilities, yielding the best performance of all tests, achieving $Q_0$ of $3.5 \times 10^{10}$ at 17.5 MV/m, an unprecedented result in multi-cell medium-beta elliptical cavities. Consistent with [16,26] we find that the 2/6 recipe has higher $R_{BCS}$ compared to the 2/0 recipe, and we therefore are inclined to consider 2/0 N-doping to be a stronger candidate for further optimization than 2/6 N-doping. At low background magnetic field levels, 2/0 doping has the lowest $R_0$ of all N-doped cavities, and the lowest $R_{BCS}$ of all recipes. Table 2 summarizes these results, in ranked order of $Q_0$ achieved at the FRIB400 operating gradient of 17.5 MV/m.

Table 2: Summary of RF results

| Treatment | $Q_0$ at 17.5 MV/m | Max Gradient | $Q_0$ at Max Gradient |
|---|---|---|---|
| **2/0 N-doping** | $3.5 \times 10^{10}$ | 22 MV/m (quench) | $2.8 \times 10^{10}$ |
| **300°C 3-h furnace baking** | $2.8 \times 10^{10}$ | 22 MV/m (quench) | $2.9 \times 10^{10}$ |
| **EP (re-baseline)** | $2.7 \times 10^{10}$ | 25.5 MV/m (admin limit reached) | $1.2 \times 10^{10}$ |
| **2/6 N-doping** | $2.6 \times 10^{10}$ | 29.5 MV/m (quench) | $1.4 \times 10^{10}$ |
| **EP (1st)** | $2.3 \times 10^{10}$ | 26 MV/m (amplifier limited) | $1.2 \times 10^{10}$ |

## IV. DISCUSSION

Achieving a $Q_0$ of $3.5 \times 10^{10}$ at 17.5 MV/m with N-doping in a $\beta_{opt} = 0.65$ 644 MHz 5-cell elliptical cavity indicates significant potential for advancement in further optimizing N-doping for the FRIB400 cavities. This cavity experienced quench at 22 MV/m $E_{acc}$, which equates to a peak B-field of approximately 97 mT. As expected, the anti-Q slope does not appear in the 650 MHz frequency range. The temperature-dependent surface resistance, $R_{BCS}$, increases consistently with field in both the 2/6 and 2/0 doped cavities, though $R_{BCS}$ remains overall lower than in the EP baselines. The 2/0 recipe shows better $R_{BCS}$ reduction than 2/6, with relatively similar medium-field Q-slopes between all the trials.

N-doping also functions as a rigorous test of this electrode design, and other EP parameter choices. Lossy surface niobium-nitrides (NbN) from the doping process must be removed with a very finely controlled light, post-doping EP of 7 μm [16].

All EP-baseline trials exceeded FRIB400 minimum $Q_0$, and no anomalous early quenches occurred, however the increase in cavity $Q_0$ after N-doping was not necessarily as high as might have been expected at 650 MHz based on high-β 650 MHz cavity results [16]. Recent work by V. Chouhan et al. [31] suggests further optimization of EP in these large cavities is possible. In particular, this work finds that the combination of a low EP cavity surface temperature with a "long" doughnut comparable to the one used in the FRIB400 cavities [7] was only moderately successful. Such an EP treatment sits on the very edge of "plateau" or polishing region of the current-voltage (I-V) curve [31]. Increasing the polishing voltage from 18V, or reducing polishing temperatures would push the treatment deeper into the plateau area, which we expect would improve upon the results presented here.

The residual resistance, $R_0$, is the strongest performance-limiting factor in N-doped cavities. A primary contributing factor to these temperature-independent rf losses is the presence of trapped background magnetic flux remaining in the cavity bulk after the superconducting transition [32]. The cavity's sensitivity to trapped flux, ($S$, in units of nΩ/mG of trapped flux) is a non-monotonic function of mean free path ($l$) at a given operating frequency, with peak sensitivity corresponds to the region of typical $l$ for N-doped cavities [33]. The most straightforward way to address increased $S$ is to minimize trapped flux.

Broadly, two options exist for minimizing the amount of trapped flux in cavities. 1) improvement of magnetic hygiene within the cryomodule to reduce the amount of background magnetic field available for trapping and 2) implementation of high thermal gradient ("fast-cooldown") to encourage flux expulsion. While 2) draws much of the technical focus, improving 1) is the more cost-effective strategy for improving $R_0$. To this end, a "slow-cooldown" under the magnetic shielding conditions used in this work is planned. If cavity performance is not significantly altered, it would motivate *in-situ* magnetic shielding techniques over investment in fast-cooldown capabilities.

Preliminary flux mitigation studies [34] have also indicated that 3 h 900°C annealing can improve the flux expulsion in some niobium cavities. However, in the case of these large, 5-cell cavities with steep sidewalls, significant concern about the mechanical integrity of baking these cavities is warranted, and multiple niobium material studies are underway to determine the effect

900°C annealing has on yield strength. Depending on project specifications for Lorentz force detuning, microphonics, maximum cryomodule loads, etc., the 900°C treatment may or may not be suitable for the given geometry of the $\beta_{opt}$ = 0.65 644 MHz cavities. Different avenues may need to be explored either to improve cavity stiffness (e.g., double stiffening rings) or to minimize background flux through improved magnetic hygene.

The furnace baking trials did not yield an anti-Q slope, and were out-performed by the 2/0 n-doping trial. Though the $R_0$ of the furnace baking trial was beneath that of the n-doped cavities, $R_{BCS}$ was elevated compared to the n-doped cavities. Even so, value of a simplified rf surface remains motivating, particularly given the successes of the single-cell $\beta$ = 1 650 MHz demonstrated in [8,9]. Recent work [35] in 1.3 GHz *in-situ* mid-T baked cavities suggests features of the $R_{BCS}$ field-dependence may be a function of oxygen diffusion depth, implying that it may be fruitful to explore different temperatures, and thus diffusion depths, at the 650 MHz operating frequency. As the work stands, the low $R_0$ makes furnace baking attractive for cavities operated at low T, where the effect of $R_{BCS}$ is negligible.

## V.    CONCLUSIONS

In order to identify the most promising high-$Q_0$ rf surface processing techniques to optimize for the novel FRIB400 upgrade $\beta_{opt}$ = 0.65 644 MHz 5-cell elliptical SC niobium rf cavities, we applied two N-doping recipes (2/6, 2/0), and one mid-T recipe, 300°C baking for 3 hours, to two prototype cavities. We find that 2/0 doping, with 7 μm post-doping EP removal tested in less than 1 mG background magnetic field delivered the highest $Q_0$ of 3.5 x $10^{10}$ at 17.5 MV/m, and experienced quench at 22 MV/m, where $Q_0$ was 2.8 x $10^{10}$. In this case, $R_{BCS}$ is reduced to about 2 nΩ, the lowest of all tested cavities. $R_0$ in less than one mG background magnetic field was limited to about 3.5 nΩ. Mid-T baking resulted in a $Q_0$ of 2.8 x $10^{10}$, with 3.4 nΩ of temperature-dependent surface resistance and 3.4 nΩ of temperature-independent surface resistance. We anticipate improvement on these results in the near future, after fuller exploration and optimization of the EP parameter space. Despite a lower level of performance in this paper, furnace baking is a promising new technique and successes elsewhere suggest that further optimizaitos can be made in the FRIB400 cavities, again with improved EP. Currently as a method for reducing $R_0$, furnace baking becomes more optimal at lower operating temperatures.

## ACKNOWLEDGEMENTS

This work is supported by Michigan State University and US Department of Energy, Office of Science, High Energy Physics Award No. DE-SC0020371. This work is additionally supported by the US Department of Energy, Office of Science, High Energy Physics under Cooperative Agreement Award No. DE-SC0018362. This material based upon work supported by the U.S. Department of Energy, Office of Science, Office of Nuclear Physics, and used resources of the Facility for Rare Isotope Beams (FRIB), which is a DoE Office of Science User Facility, under award Number DE-SC0000661.